\documentclass[copyright,creativecommons]{eptcs}
\providecommand{\event}{F-IDE 2019} 
\usepackage{breakurl}             
\usepackage{underscore}           
\usepackage{subfig}
\usepackage{cite}
\usepackage{graphicx}
\usepackage{tikz}
\usepackage{listings}
\usepackage{subfig}
\usepackage{amsmath}

\usepackage{amsthm}
\usepackage[labelfont=bf]{caption}
\usepackage{amssymb}
\usepackage{xspace}
\usepackage{todonotes}
\usepackage{enumerate}
\usepackage{hyperref}
\colorlet{keywordcolor}{blue!50!black}
\colorlet{commentcolor}{green!60!black}
\colorlet{typecolor}{violet}
\newcommand{\sourcefont}{\ttfamily\small}
\newcommand{\commentfont}{\slshape\rmfamily\color{commentcolor}}

\lstdefinelanguage{ABS}{
        keywords={assert,this,new,data,type,def,case,of,local,class,interface,
        extends,implements,if,then,else,await,get,Fut,return,skip,while,module,
        import,export,from,to,suspend,delta,adds,modifies,removes,original,productline,
        features,core,corefeatures,optionalfeatures,after,when,product,hasAttribute,
        hasMethod,hasField,hasInterface,uses,root,extension,group,allof,oneof,require,
        stateupdate,objectupdate,classupdate,
        exclude,original,ifin,ifout,opt,null,
        newgroup,data,thiscomp,in,joins,leaves,subtypeOf,wait,acquire,except,as,component,Pre,Abs
        },
        keywordstyle=\color{keywordcolor}\bf\sffamily,
        morekeywords=[2]{Unit, Int, Bool, Rat, List, Set, Pair, Fut, Maybe, String, Triple, Either, Map},
        keywordstyle=[2]\color{typecolor},
        sensitive=true,
        comment=[l]{//},
        morecomment=[s]{/*}{*/},
        morestring=[b]"
}

\lstdefinelanguage[v9]{Java}[]{Java}{
        morekeywords={module,requires,provides,uses,with,to,exports}
}
\lstdefinelanguage[ContextJ]{Java}[]{Java}{
        morekeywords={layer,with,without,proceed,before,after}
}
\lstdefinelanguage[FOP]{Java}[]{Java}{
        morekeywords={refines,original,Super}
}
\lstdefinelanguage[JastAdd]{Java}[]{Java}{
        morekeywords={aspect,syn,inh,lazy}
}

\lstdefinestyle{code}{
        basicstyle=\sourcefont\upshape,
        keywordstyle=\color{keywordcolor}\bf\sffamily,
        commentstyle=\commentfont,
        columns=fullflexible,
        mathescape=true,
        escapechar={\#},
        keepspaces=true,
        showstringspaces=false,
        inputencoding=utf8,
        extendedchars,
        aboveskip=8pt, 
        numbers=left,
        stepnumber=1, 
        numberstyle=\ttfamily\scriptsize\color{gray},
        numbersep=4pt,
        xleftmargin=1.5em,
        xrightmargin=1.5em,
        framexleftmargin=1.2em,
        framexrightmargin=1em,
        framextopmargin=0.5ex,
        breaklines=true,
        breakindent=3pt,
}

\lstdefinestyle{abs}{
        style=code,
        language=ABS,
}
\lstdefinestyle{java}{
        style=code,
            language=Java
}
\lstdefinestyle{java9}{
        style=code,
            language=[v9]Java
}
\lstdefinestyle{aspectj}{
        style=code,
        language=[AspectJ]Java
}
\lstdefinestyle{jastadd}{
        style=code,
        language=[JastAdd]Java
}
\lstdefinestyle{contextj}{
        style=code,
        language=[ContextJ]Java
}
\lstdefinestyle{FOP}{
        style=code,
        language=[FOP]Java
}
\lstdefinestyle{scala}{
        style=code,
        language=Scala,
        morekeywords={self}
}

\newcommand{\code}[2][]{\lstinline[style=code,basicstyle=\ttfamily\upshape,#1]{#2}}

\newcommand{\abs}[2][]{\code[style=abs,#1]{#2}}

\lstnewenvironment{srccode}[1][]{
        \minipage{\linewidth}
        \lstset{style=code,
        backgroundcolor=\color{white},
        rulecolor=\color{gray!50},
        frame=tblr,
        captionpos=b,
        #1}
}{\endminipage}

\lstnewenvironment{abscode}[1][]{
        \minipage{1\linewidth}
        \lstset{style=abs,
        backgroundcolor=\color{white},
        rulecolor=\color{gray!50},
        frame=tblr,
        captionpos=b,
        #1}
}{\endminipage}

\lstnewenvironment{javacode}[1][]{
        \minipage{\linewidth}
        \lstset{style=java,
        backgroundcolor=\color{gray!5},
        rulecolor=\color{gray!50},
        frame=tblr,
        captionpos=b,
        #1}
}{\endminipage}

\lstnewenvironment{jastaddcode}[1][]{
        \minipage{\linewidth}
        \lstset{style=jastadd,
        backgroundcolor=\color{gray!5},
        rulecolor=\color{gray!50},
        frame=tblr,
        captionpos=b,
        #1}
}{\endminipage}

\newcommand{\visbar}{\texttt{VisualisierbaR}\xspace}
\newcommand{\formbar}{\texttt{FormbaR}\xspace}
\newcommand{\ask}[1]{``\textit{#1}''\xspace}

\newcommand{\COMMENT}[1]{}
\newcommand{\FULL}[2]{#1}

\newtheorem*{example*}{Example}
\newtheorem{definition}{Definition}
\newtheorem*{definition*}{Definition}

\title{Tool~Support for Validation of Formal~System~Models:\\ Interactive~Visualization and Requirements~Traceability}

\author{
Eduard Kamburjan
\institute{Department of Computer Science\\ Technische Universit{\"a}t Darmstadt, Germany\\ \texttt{kamburjan@cs.tu-darmstadt.de}}
\and
Jonas Stromberg
\institute{Department of Computer Science\\ Technische Universit{\"a}t Darmstadt, Germany\\ \texttt{jonas.stromberg@stud.tu-darmstadt.de}}
}

\def\titlerunning{Tool~Support for Validation of Formal~System~Models}
\def\authorrunning{E. Kamburjan \& J. Stromberg}

\begin{document}
\maketitle

\begin{abstract}
Development processes in various engineering disciplines are incorporating formal models to ensure
safety properties of critical systems. The use of these formal models requires to reason about their
adequacy, i.e., to validate that a model mirrors the structure of the system sufficiently that properties
established for the model indeed carry over to the real system. Model validation itself is non-formal,
as adequacy is not a formal (i.e., mathematical) property. Instead it must be carried out by the modeler
to justify the modeling to the certification agency or other stakeholders. In this paper we argue that
model validation can be seen as a special form of requirements engineering, and that interactive visualization and concepts from
requirements traceability can help to advance tool support for formal modeling by lowering the cognitive burden needed for validation. We present the
VisualisierbaR tool, which supports the formal modeling of railway operations and describe how
it uses interactive visualization and requirements traceability concepts to validate a formal model.

\end{abstract}

\section{Introduction}
The importance of formal methods for safety-critical systems has long been recognized in many engineering disciplines and is demanded or recommended by certification authorities in, e.g., railway engineering~\cite{en} and avionic~\cite{do} industries. 
Recently, with the increasing integration of computational parts into devices, digital twins~\cite{twin} and co-simulation~\cite{cosim} are used to develop new products and prototype changes.
One important class of formal methods in this area is formal system modeling.

Under formal system modeling we understand the development of a formal system model of a real system or of a design of a planned system (short: target system) that mirrors the structure and behavior of the target system
sufficiently to prototype~\cite{proto} and/or evaluate changes\footnote{We contrast formal system modeling with Model-Driven-Development approaches, where the model becomes the final system through refinement.}~\cite{cosim}.
Digital twins are a variant of this, which are integrated into the target system. 
Nonetheless, digital twins are based on a subsystem whose structure they must mirror as close as possible and face the same challenges for validation.
Formal system modeling requires model validation to ensure that properties established for the formal model hold for the target system:
While verification ensures that the model \emph{behaves correctly}, validation ensures that the \emph{correct thing} was modeled.
Model validation itself is not formal (in the sense that it is not a mathematical property), it is an informal process to argue for the adequacy of the model and bridges between the intention of the developer and the realized model.
Validation is required to convince safety assessors ,such as certification agencies, that formal proofs have value in the certification process and other stakeholders that the prototypes developed in this model save development time for the target system.

Our main observation in modeling projects with industry partners is that certain stages of formal modeling can be seen as a specific form of requirements engineering.
\begin{itemize}
\item
\emph{Requirements elicitation is model scoping}. Both these processes turn the \emph{implicit} knowledge and assumptions of the user about the domain into an \emph{explicit} representation. 
More importantly, they also decide on the aspects of the domain that are not needed for a specific model/project.
\item 
\emph{Requirements traceability is model validation}. Both tasks relate parts of the formal system model to the target system, which are the two main artifacts from requirements engineering view.
Instead of tracing a requirement to the point where it is realized, one traces an aspect to the point where it is modeled.
In reverse, instead of tracing backwards what requirement a part of the implementation is realizing, one traces what aspect a model-part realizes.
\end{itemize}
However, formal modeling poses challenges that prevent the straightforward adaption of, e.g., Software Engineering practices.
The boundary between modeling and programming is not clear~\cite{isola}, but concerning validating programs and formal system models of the described kind the main difference is that model validation of formal system models requires to validate a \emph{white-box} model (in contrast to a black-box model when testing a program) and, in particular, raises the following points:
\begin{description}
\item[Cognitive Burden.] 
Significant cognitive burden is required to judge formal system models, as formal modeling languages are not adopted by all industries and there is little training material available.
Even in industries which use formal modeling, keeping the cognitive burden low is a desired aspect of adopters of formal methods in industry~\cite{tla} and the cognitive burden of validation is higher than when designing, e.g., use cases and user stories.
\item[Validating Structure.] It is conceptually different to validate the \emph{structure} of a white-box model, than to validate the \emph{behavior} of a black-box model~\cite{barlas}.
As formal modeling aims to mirror the structure precisely enough that changes in the model have the same causal effect as their counterpart in the target system, merely describing (by, e.g., test cases) the input/output does not suffice.
This thwarts the application of behavior-centric approaches such as Behavior-Driven Development (BDD)~\cite{bdd}.
\end{description}

Another experience we make in our work with domain experts is that merely \emph{visualizing} a formal system model is not enough when using it to prototype new ideas~\cite{proto}.
Interaction with the visualization allows even quicker feedback cycles with the domain experts, as it allows them to test a specific situation for validation with little overhead to induce it into the model.
This ties in with the above point of lowering the cognitive burden to simplify validation.

These observations raise the question how, and what, techniques for requirements traceability and interactive visualization can be applied to formal modeling.
In particular, we are interested in integrating such techniques into an IDE that helps not only with model verification, but also with model validation.

We illustrate with the \visbar tool for formal modeling of railway operations how requirements traceability can be integrated into a formal methods toolkit
and describe future research directions for formal modeling languages and toolkits. 

Our main contribution is to develop tool support for model validation by intergrating requirements traceability and interactive visualization into an IDE, as well as a tool and a case study illustrating this idea.
This work is structured as follows: Sec.~\ref{sec:validate} describes model validation in railway operations, 
Sec.~\ref{sec:formbar} gives an overview over a formal model in this domain, 
Sec.~\ref{sec:impl} descirbes the implementation of the \visbar tool,
Sec.~\ref{sec:overview} describes the validation features , Sec.~\ref{sec:workflow}
gives two case studies from automatic train operations and rule prototyping and Sec.~\ref{sec:conclusion} concludes with related and future work.

\section{Validation of Railway Operation Models}\label{sec:validate}
We describe our approach using the \visbar tool developed for the \formbar model~\cite{scp} for German railway operations.
This section describes the specification of railway operations and the challenges of validating in this domain,
while the approach itself is easily generalized to other domains.
\subsection{Specification of Railway Operations}
Railway operations for German railways are not described by a single document, 
but by 
(1) legal regulations, the \emph{``Eisenbahn-Bau- und Betriebsordnung'' (Law for Operating and Building Railways)}~\cite{eba}, 
(2) public rulebooks managed by Deutsche Bahn (DB), in particular Ril.~408~\cite{ril408} and 819~\cite{ril819}, 
(3) internal rulebooks for operations, 
(4) requirements specification for technical elements, 
(5) training documentation and (6) internal announcements. 
\formbar only considers the operations of DB, but other railway companies are also bound to the same legal regulations (and to Ril.~408 when using DB infrastructure), in addition to their own internal rulebooks. 

In this environment, procedures are not described algorithmically in one place, but are described in a distributed manner. 
This makes it hard to pinpoint the exact point where the procedure is defined. 
E.g., the procedure to depart a train is partially described by Ril.~408~\cite{ril408}, 
partially by the requirements of the specific station interlocking in a station, partially by internal announcements and possibly by local exemptions (\emph{``Lokale Zus{\"a}tze''}).
Implicitly, building regulations are also referenced, as certain minimal distances are assumed to hold.

These procedures are subject to constant change and completely new procedures for automatic train operations (ATO) and ETCS level 3 are in development.
When scoping \formbar, it was decided not to model certain rules, because they are not relevant to train operations itself but, e.g., specify interactions with the passengers~\cite{scp}.
\subsection{Validating Railway Operation Models}
Validation of models of new procedures requires to track each part of the model to the document that specifies it --- legal regulations, rulebooks and technical documents are requirements and model validation entails documenting that the requirements are met.
This is especially critical if these models are planned to be used for certification.

However, contrary to engineering projects, the form of the requirements is already fixed in a form that is optimized towards other uses -- 
rulebooks are a form, which is difficult to process and which may not be changed during development.
In terms of characteristics for software requirements~\cite{IEEE830}, they are neither unambiguous, nor complete or modifiable\footnote{Arguably, they are also neither consistent nor structured by importance.}. 
Furthermore, while there is a public specification in DOORS format available for the new European Train Control System (ETCS) modes of operations, the other rulebooks are written in plain natural language structured by sections and paragraphs.
The use of technical documents that were not intended to be used as requirements for formal system models is not specific to railway engineering but is common in other fields, as formal models are mostly developed after the target system is finished.
Similarly using the requirements of the original system can be problematic, as the formal model then expresses what the system is \emph{supposed to do}, in contrast to what it really \emph{does}.
Nonetheless this can be of use, e.g., to analyze the design before implementation.

Not the complete model is directly related to requirements: some parts model basic infrastructure.
E.g., \formbar contains code for the physical behavior of the train, which is not explicit in any rulebook.
For validation it is important to carefully distinguish between basic infrastructure and other model parts, since an error in the basic infrastructure is a mistake of the modeler, while
an error in the other parts may hint towards a problem with the target system.

\subsection{Validating \formbar}
We use three techniques to validate \formbar: simulation, interactive visualization and traceability.
\begin{description}
\item[Simulation.] 
Simulation runs the model on predefined infrastructures and scenarios and checks that the behavior is the one expected by the domain expert, e.g., that after a fault on the infrastructure the train has the expected delay.
This roughly corresponds to acceptance testing for software, but does not scale for bigger scenarios, e.g., because propagation of faults is not easily specifiable or predictable.
Visualization scales better, as it is easier for a human to assess the visualized situation than to assess (or specify) the expected behavior as a trace.
\item[Interactive Visualization] Simulation is only able to detect errors in simple scenarios. A visualization tool shows the state of the whole infrastructure, e.g., the position of the train or the state of the signals.
Interactive visualization is not merely a representation of the behavior of the system. The user interacts with the model via the visualization and introduces faults or gives orders to the train.

Interaction extends the use of visualization for validation. 
First, it is easier for the domain expert to assess the adequacy of the model if larger parts of the model can be inspected easily.
Second, by interacting with the model he can explore the behavior of the model for questions arising during the validation. 
E.g., to check whether a certain combination of faults has been modeled correctly, when the interactions of faults is scattered in the model.

\item[Requirements Traceability] Simulation and interactive visualization treat the model as a black box and merely ensure that the behavior of the model corresponds to the expectations of the domain expert in a number of situations.
To ensure that the \emph{internal structure} of the model mirrors the internal structure of the domain we annotate the model and the visualization with links to the text files containing the specification.

Requirement links trace a requirement either forward (answering \ask{where is this requirement realized?}) or backward (answering \ask{what requirement does this code realize?}).
Similarly, annotations are two-directional. 
A section of a rulebook links to the code in the model that implements it (i.e., is a forward trace link) and the code links to the rulebook it implements (i.e., is a backwards trace link).
Links between model and specification are not enough, as the code may still implement a procedure that is a described in several places. 
Visualization allows us to output messages that also contain links to the specification, to connect these representation without explicitly invoking the model.
These links serve two purposes: First, they enable us to track in the visualization whether the procedure is executed correctly (i.e., according to specification). 
Second, they ensure that the visualization, which is an additional abstraction layer/artifact (additionally to rulebooks and formal model) is integrated into the validation of the formal model.
\end{description}

\section{ABS and the \texttt{FormbaR} Model}\label{sec:formbar}
In this section we give a short overview over the Abstract Behavioral Specification (ABS) language~\cite{abs} and the \formbar~\cite{scp} model of railway operations. 
For brevity's sake, we only introduce ABS and \formbar as far as needed to explain \visbar; an introduction to ABS can be found in~\cite{abs}, an extended description of \formbar in~\cite{scp}.

\subsection{Abstract Behavioral Specification}
ABS is a modeling language, developed for the modeling, simulation and analysis of distributed systems. 
ABS models are executable, yet it is not a programming language in this context: its foremost use is to mirror the structure of the target system, not its computational results.
Its conceptual closeness to programming languages, however, allows us to demonstrate the use of requirements traces more succinctly.
Most constructs of ABS are standard and its syntax is based on Java, with additional statements for concurrency.
We introduce the data, communication and time models of ABS. 

\paragraph{Data and Communication Model.}
ABS models data and behavior in two sublanguages.
Data, and operations on the data, is modeled in a functional sublanguage based on abstract data types (ADTs). As an example, the following defines an ADT modeling the state of a (logical) signal:
\\\noindent\begin{abscode}[numbers=none]
data State = GO | HALT | SLOW | INVALID;
\end{abscode}

Behavior and communication is modeled in an object-oriented language on top of the functional sublanguage. 
The object model uses classes and interfaces and is based on Java, but all fields are object-private. 
Additionally, traits may be used to add methods to a class.
The following class models a Zs10 auxiliary signal (end of speed limitation).
\\\noindent\begin{abscode}
[Concept:"Zs10"] class Zs10(Edge track, String name) {
  uses NoSig adds NoBack adds Nameable;
  List<Trans> trigFront(Train t, Edge e){
     Information info = NoInfo;
     if( e == track ) 
       info = AreaEnd(-1, False, null);
     return list[Pass(info)];
  }
}
\end{abscode}

The class \abs{Zs10} has two fields (\abs{track} and \abs{name}) and uses three traits (\abs{NoSig}, \abs{NoBack}, \abs{Nameable}). It has one additional method, that transmits \abs{AreaEnd} if the train passes the Zs10 
auxiliary signal from the direction where it is visible. The types \abs{Trans} for transmissions and \abs{Information} for transmitted information are ADTs.

\COMMENT{\paragraph{Concurrency Model.}
ABS has a preemption-free concurrency model: Once a process is started, other processes on the same object may only run if the active process actively suspends itself or terminates.
To suspend itself, the process uses the \abs{await} statement. There are multiple ways to use this statement, one of them is to synchronize on some boolean condition. 
E.g., \abs{await this.i >= 0} suspends the process, until the guard expression \abs{this.i >= 0} holds. A suspended process is inactive until its guards holds. Each time a 
process is suspended or terminates, a waiting process (which can continue) is chosen non-deterministically for reactivation. Expressions are side-effect free, so reevaluating the guards does not change the object state.

Interactions between multiple objects are realized via asynchronous method calls. Such a call \abs{f = o!m()} to a method \abs{m} returns a \emph{future} to the caller.
The caller continues execution, without waiting for the called process to finish. If the result of the called method is needed, the caller can synchronize on the future.
Synchronization is either non-blocking via \abs{await f?}, which suspends a process until the process of the future stored in \abs{f} has terminated or blocking via the statement \abs{v = f.get}, which reads the result of the 
process of the future stored in \abs{f}, once it has terminated. Until then, the object is blocked for other processes. 
}
\paragraph{Simulation, Time and Model API.}
\formbar uses Timed ABS~\cite{time}, which extends ABS with explicit operations on time. 
The statement \abs{await duration(x,y)} suspends the current process for at least \abs{x} to \abs{y} simulation time steps.
At runtime, the shortest possible time is chosen.
In \formbar, a simulation time step corresponds to one second.

ABS can be compiled into, among other languages, Erlang and then be executed. 
The compiled executable contains a runtime environment in Erlang, that implements the above concurrency model and keeps track of the symbolic time --- 
the global symbolic clock is only advanced if every object is waiting for time to pass. 
The clock is then advanced by the minimal time that unlocks some object.

\subsection{The \formbar model of Railway Operations}
The \formbar model is centered around the notion of \emph{points of information flow}, which are the basic infrastructure of the formal model and are not specified by the rulebooks.
\begin{definition}
A \emph{Point of Information Flow} (PIF) is an object at a fixed position on a track, where one of the following applies:
\begin{itemize}
\item It is an infrastructure element transmitting information to trains (e.g., balises)
\item It is in some critical distance to an infrastructure element (e.g., the point where the presignal is seen at the latest)
\item It is an infrastructure element, which receives information from trains 
\end{itemize}
\end{definition}
PIFs allow one to discretize the infrastructure from an \emph{operational} perspective, as the physical behavior of the train can be interpolated between two PIFs.
\formbar is also able to handle state changes of trains between two PIFs, e.g., because of orders or if the train comes to a halt before a signal.
However, for the most time during simulation, the train behaviors must only be adjusted at PIFs and simulation is thus less time-consuming.

\subsection{Infrastructure}
The infrastructure model is based on a graph, where the nodes form the base for a four-layer model of the infrastructure.
At each node of the graph, PIFs may exist and the edge has the length of the track in between two nodes.

This topological graph forms layer 1 and contains all information about physical distances. Layer 2 is a set of physical elements assigned to a node, e.g., presignals, main signals, etc.
Furthermore, layer 2 is the view of the train driver on the infrastructure, who has to react to these elements.
Layer 3 consists of \emph{logical elements}. A logical element is a set of physical elements, which share state or interface to the interlocking system.
This layer is the view of the train dispatcher on the infrastructure, as it is not possible to, e.g., change the state of the main signal without changing the state of the presignal.
A physical element may be assigned to multiple logical elements (e.g., a presignal may belong to multiple logical signals) or none (e.g., a buffer stop).

\begin{figure}
\centering\includegraphics[scale=0.4]{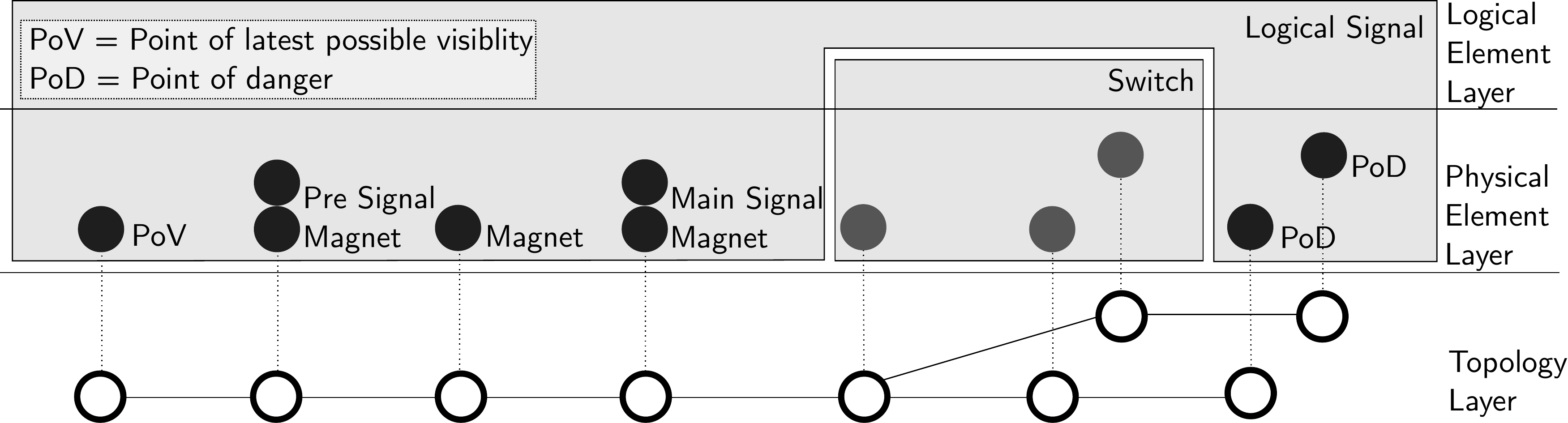}
\caption{The lower three layers of a station entry, with a logical entry signal and a switch. (From~\cite{scp})}
\label{fig:layer}
\end{figure}

Fig.~\ref{fig:layer} shows the entry to a train station. The black elements constitute one logical signal, the entry signal of the station: 
The point of visibility, where the presignal is seen at the latest, the presignal itself, the main signal, three magnets of the automatic train protection system PZB and two point of danger which are covered by the signal (e.g., axle counters).
\visbar has basic CAD features to create and manipulate infrastructure.

\subsection{Communication of Stations and Trains}
Stations manage a set of logical elements and communicate only with their logical elements and adjacent stations.
Trains only communicate with the lowest layer of the infrastructure, the graph. The nodes relay all the transmissions from the physical elements on them to the train.
Trains and stations communicate directly only via orders and only in case of faults, not during normal operations.

\section{Implementation and Interaction}\label{sec:impl}
In the following sections we describe \visbar, an IDE that implements the principles of the previous section and illustrates prototyping of railway operation procedures.
The interface has three components: an ABS IDE for the model, a PDF viewer for the rulebooks and a visualization.
\visbar is designed for a multi-monitor working place, due to the space requirements of the visualization of the simulation.
First, we describe the architecture of the implementation and the possibilities to interact with the model in \visbar.

\subsection{Implementation}
\begin{figure}
\centering\includegraphics[scale=0.6]{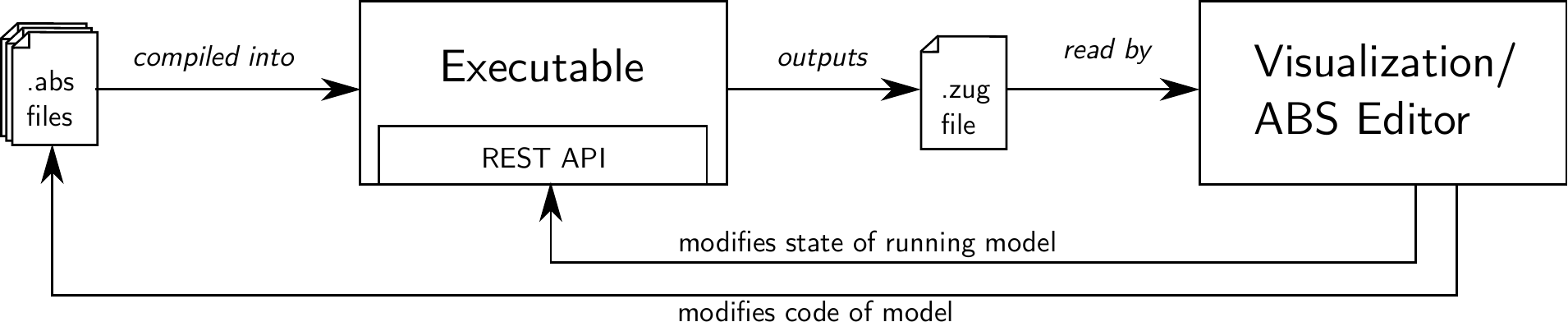}
\caption{Overview over the components in the implementation}
\label{fig:impl}
\end{figure}
To start the simulation, the following workflow is implemented:
The ABS compiler first generates Erlang code and then compiles Erlang to an executable file. The executable outputs a \texttt{.zug} file.
This file contains a list of all \formbar events that are needed for the visualization and acts as the interface between visualization and the model.
It also allows to replay an execution without having the ABS code. The format is a list of events, e.g., the following is a main signal with the internal Erlang identity \texttt{TrackElements.HauptSignalImpl:<0.581.0>} changing its state to ``Go''(Fahrt) at $459/8s$.
\begin{lstlisting}
CH;TrackElements.HauptSignalImpl:<0.581.0>;FAHRT;459/8
\end{lstlisting}

Additionally, the executable contains a web server running a RESTful API~\cite{arbab} to query the object state, call methods from the outside and to limit the clock. 
By limiting the clock, it is possible to start the executable, read (and visualize) the output up to a certain time step and then interact with the model
by calling methods via the RESTful API. Afterwards, one may resume the execution for some fixed time span by increasing the limit of the clock.
\visbar requires that the ABS project consisting from \texttt{*.abs} files and a scenario are selected.
It automatically compiles the model and starts the executable. 
We use two kinds of interactions from \visbar, which are illustrated by the two cycles in Fig.~\ref{fig:impl}.

\subsection{Interaction with Running Model} \label{sssec:Interaction with Running Model}
Interaction with the running model is the inner cycle in Fig.~\ref{fig:impl}.
As described, the web server allows us to interact with an already running model by invoking exposed methods.
In the visualization, each physical element displays the interactions.
An interaction is a method which is exposed by an \abs{[HTTPCallable]} annotation in the interface of the class. 

The RESTful API is used to read the list of exposed methods and allows to easily add methods as new interactions (at compile time).
When the simulation is halted, these methods may be called to change the current state and alter the following steps in the simulation.
In principle, the model must not be halted for the interactions, however \formbar does not adjust simulation and wall time, to interact on a precise point in time, one thus needs to simulate up to this point, interact, and continue simulation.
These interactions are available for trains and physical elements, which may however propagate the interaction to their currently responsible train station, resp. logical element. 

\subsection{Interaction with ABS Code}
Interaction with the ABS code is the outer cycle in Fig.~\ref{fig:impl}.
\visbar contains an IDE for ABS, which allows to run the simulation, visualize it and then directly modify the ABS model. 
This allows visual \emph{debugging} of railway operations, where certain situations can be modeled as the infrastructure and then directly checked whether the new (or modified) procedure behaves as intended.
After modifying the code, the model is recompiled and re-executed. 

To support this interaction, it is necessary to provide a way to link the visualization with two parts of the ABS code: First, the infrastructure that is currently active and second, the part of the procedure that is executed.
This connection falls under contextualization, which does not only provide the context of rulebooks for the ABS model, but also the context of the ABS model for the visualization.
When modifying the infrastructure, the complete initialization block of the scenario if generated anew.

\section{Using \visbar for Validation}\label{sec:overview}
In this section we describe the use of \visbar for validation. First, we describe the visualization.

\subsection{Visualization}
\paragraph{Modes.}
\visbar can be started in three modes. 
If \visbar is started in \emph{Visualize/Edit} mode, then the code of the ABS model and the visualization of the simulation are shown. 
The scenario can be edited, the simulation can be rerun and the contextual documents can be displayed.
If \visbar is started in \emph{Interactive} mode, then the visualization of the simulation is shown. 
The simulation, however, is not run yet. Instead the visualization offers the opportunity to either interact with the halted simulation (e.g., to inject faults by breaking signals or to give orders to a train) or 
decide to continue for a certain time frame. The ABS model and contextual documents can be displayed, but the scenario can not be edited. 
In these two modes, the root directory of the ABS model and the chosen scenario have to be selected.
Finally, the \emph{Replay} mode allows to visualize \texttt{.zug} files without ABS model.

\begin{figure}[bth]
\begin{center}
\centering\includegraphics[scale=0.45]{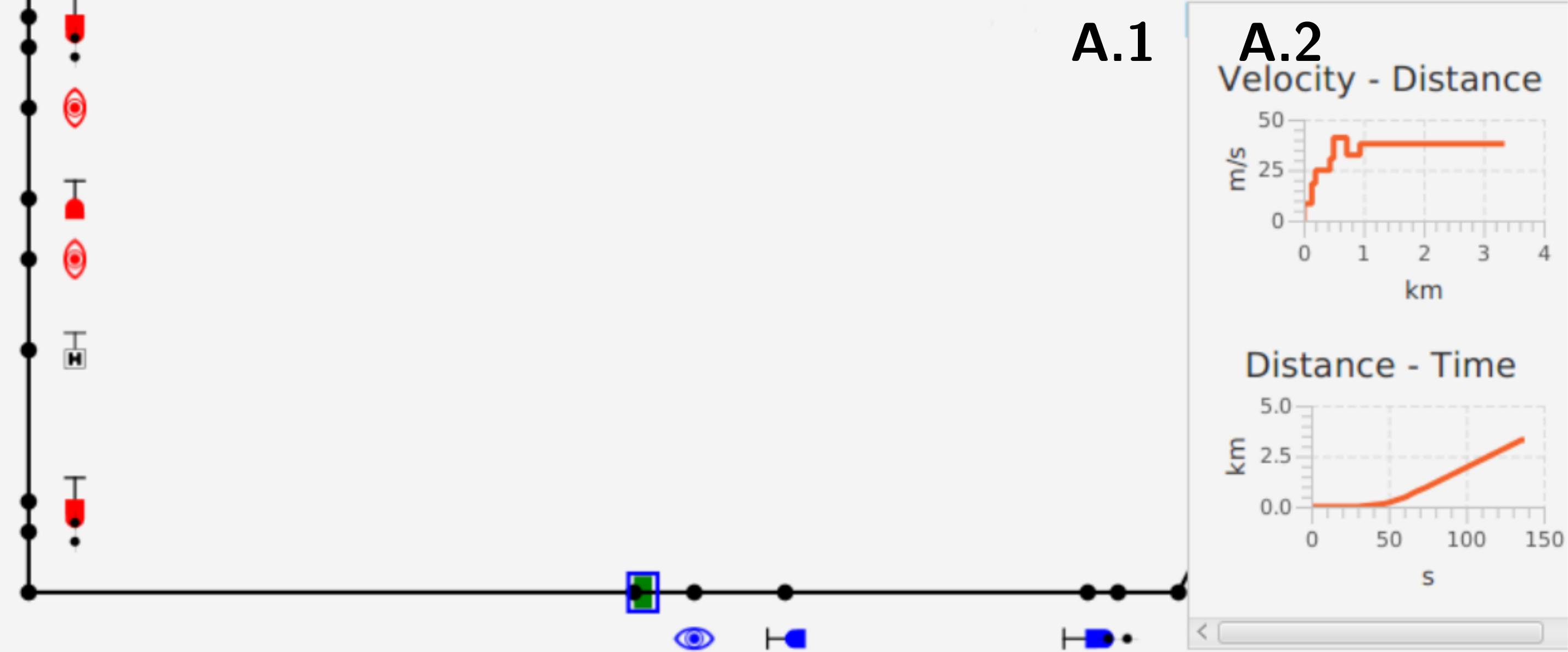}
\end{center}
\caption{Window A: Visualization. The green train is about to enter the station on the bottom right, the entry signal shows ``Go''. Window A.2 shows the $v$-$t$ diagram and other information about the train.}
\label{fig:showB}
\end{figure}

A detail of the visualization of railway operations is shown in Fig.~\ref{fig:showB}. This window offers, beyond visualization itself, the ability to interact with the model and 
can be used as an editor, that offers standard computer aided design (CAD) features: adding, editing, copying and deleting nodes, edges and physical elements.
It is also possible to manage logical elements. 
If the ABS model is changed, the model is recompiled and the visualization shows the rerun scenario. 

Window A.2 shows the details of a train or element, if one is selected. 
For physical elements, properties such as position and its logical element are displayed.
For trains, additionally to the properties, the $v$-$t$ and other graphs are shown. 
In interactive mode, A.2 also contains the possible interactions with the selected element or train.
Optionally, a window with the list of \texttt{FormbaR}-events can be opened.
There are 3 possibilities to advance the simulation: 
\begin{itemize}
\item
by manually selecting a point in time or event, 
\item
by traversing the list of events automatically at a fixed rate events/second,
\item
by traversing the list of events automatically at a fixed speed. In this case the position of trains between two nodes is interpolated.
\end{itemize}
In any case it is possible to go back in time and review a part of the simulation. It is however only possible to interact with the current (i.e., newest) state.

Interaction is realized through a RESTful API~\cite{arbab} embedded in the ABS executable that allows to call methods from the outside and to limit the clock.
The executable outputs \texttt{*.zug} files, which are read by the visualization.
Reading these files allows one to replay an execution without needing the ABS code, which simplifies sharing.

\subsection{Requirements Traces}

\begin{figure}[bth]
\begin{center}
\includegraphics[scale=0.45]{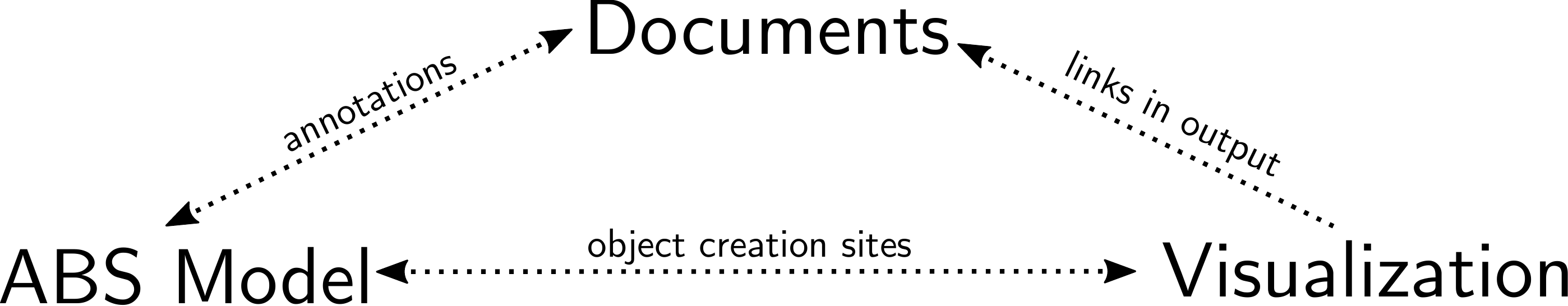}
\end{center}
\caption{Structure of trace links in \texttt{VisualisierbaR}.} 
\label{fig:struct}
\end{figure}

The links between the components are illustrated in Fig.~\ref{fig:struct}:
\begin{description}
\item[Documents to ABS]
By selecting a part of the rulebook that is linked from the code, the linking code, i.e. the annotated element, can be highlighted.
\item[ABS to Visualization]
Objects are highlighted in the visualization, if their object creation site (their \abs{new} expression) is selected in the ABS editor.
\item[Visualization to Documents]
The visualization allows us to show explaining text for the simulation. These messages may contain the annotations to the rulebooks.
\item[Visualization to ABS]
The object creation sites of elements selected in the visualization are highlighted in the editor.
\item[ABS to Documents]
The ABS code allows us to directly link to the rulebooks from the code via annotations. The relevant part of the rulebook is then highlighted.
\end{description}
The first two links implement forward tracing, the others implement backward tracing.
The links between ABS and documents support $n$-$m$ relations -- if a part of the document is modeled in several points of the code, a window allows to select one.
The following sections illustrate the trace links in more detail. 

\paragraph{Visualization.}
The visualization provides context in two ways: (1) when selecting an element, window A highlights the statement responsible for the creation of this element. This link is used to trace an element in the visualization to a point in the model.
(2) Additionally, special \texttt{MSG}-events in the \texttt{*.zug} files are supported: These events are shown as pop-ups and visualize non-visible state changes (e.g., message exchange between train and station).
An \texttt{MSG}-event may contain annotations to link to contextual documents. This link traces a point in the execution of a model to a rulebook/requirement. 

\begin{figure}[bth]
\begin{center}
\includegraphics[scale=0.45]{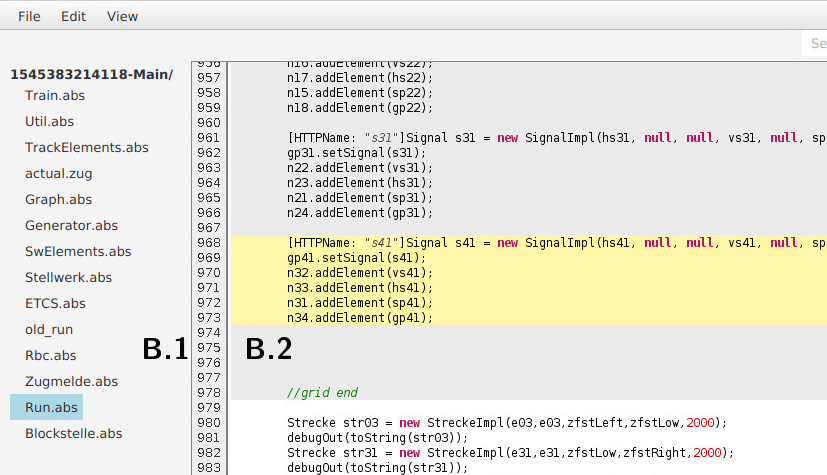}
\end{center}
\caption{Excerpt of Window B: ABS editor. The different backgrounds visualize infrastructure selected in window A.} 
\label{fig:showA}
\end{figure}

\paragraph{ABS Model.}
The ABS model is shown in window B in Fig.~\ref{fig:showA}. 
ABS is a modeling language with a Java-style syntax and is presented similar to mainstream programming languages in the IDE.

In window B a file browser shows the different code files (B.1), while the main part (B.2) allows one to view and manipulate a single ABS file.
B.2 offers standard IDE features like syntax highlighting or jumping to definitions. 

Specific to \visbar are two features that provide contextualization:
\begin{enumerate}[I]
\item In the scenario setup in file \texttt{Run.abs}, each created element can be clicked on and is then highlighted in the visualization (window A).
This link is used to trace a part of the model to the visualization. 
\item Each class and method can be annotated with \abs{[Document:Y]}, where \abs{Y} is a rulebook identifier (e.g., ``Ril. 408.0615'') the name of a concept or a keyword, (e.g., ``Main Signal'').
A click on such annotations highlights the document part in the document window (window C) marked with this identifier (in case of \abs{[Document:Y]} or a window that lists all document parts responsible for the keyword (in case of \abs{[Concept:X]}).
This link is used to trace a part of the model to the rulebook/requirement. 
If a concept is linked to multiple parts of the rulebooks, the user can select one of them. 
The mapping between annotated concept and rulebook sections are manually managed in a \abs{.csv} file that allows $n$-$m$ relations and is a variant of a requirement matrix between code and rulebooks.
\end{enumerate}

The IDE offers a way to modify the code and recompile. After recompilation, window B.2 is split into two panes, where the left shows the current code and the right the code of the model before compilation to simplify tracking of changes.

\paragraph{Documents.}
\begin{figure}[bth]
\begin{center}
\includegraphics[scale=0.3]{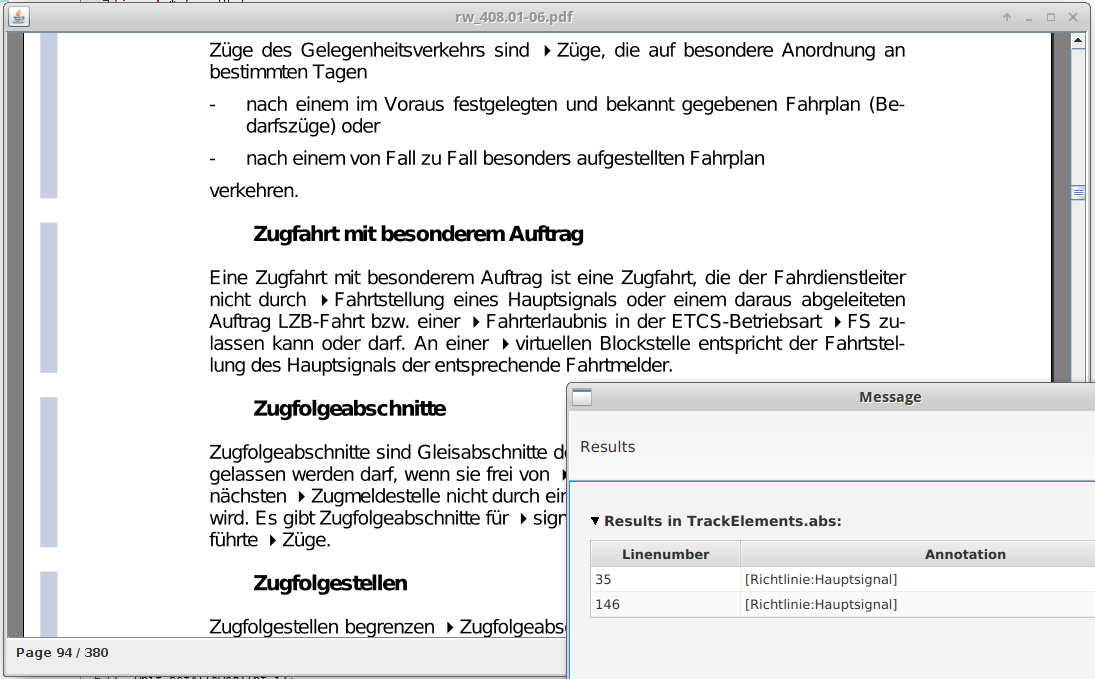}
\end{center}
\caption{Window C: PDF viewer}
\label{fig:showC}
\end{figure}
Window C is a PDF viewer which highlights parts of the document if referenced from the other components.
It provides context in two ways. When a part of the document is clicked on and this part of the document is referenced from the model, a list of all annotation referring to it is shown. This list then highlights the annotation in window B.2.
This link is used to trace a requirement to the model. 

Fig.~\ref{fig:showC} shows the PDF viewer. The bars on the left mark the referenced parts of the documents. The window on the right is displayed when a part of the document is clicked on and lists all references in the ABS code to it.
A list of all references can be shown in the ABS editor.

\section{Validation Case Studies for ATO and Prototyping}\label{sec:workflow}

We give two examples how \visbar can be used in the workflow of rulebook authors.
The first example is from the on-going development of new procedures for autonomous train operations~\cite{Bilal1}, where \visbar was applied to check that the new rules correctly interact with the old rules for non-autonomous train operations.
The second example models the change of a rule. Followine real world changes~\cite{pachlBahn}, we model the effects on delays, depending on whether the first train after a fault has occurred drives on sight or not.
\subsection{Validating ATO Procedures}
We give an example where \visbar is used in the current development of a system to handle faults during autonomous train operations (ATO) with grade of autonomy (GoA) 4~\cite{Bilal1} 
to analyze how the additional checks needed for ATO interact with the operational rules. 
In this case, the requirements are the developed procedures for ATO GoA 4 and their correct interactions with the original rules.

The investigated scenario was an obstacle in front of a signal, where
the autonomous train adheres to the rules specific to GoA 4 (detecting the obstacle and waiting for it to disappear) as well as to the rules for general operations (responding to the signal).
The model has to realize both rulebooks. 

The decisions necessary for ATO are annotated with links to the ATO documentation\COMMENT{In this case a UML activity diagram. We annotated statements with links to activities} and the already existing model for general operations with links to the rulebooks.
To connect the model with the visualization, we added messages and used the following scenario:
A train is driving towards a signal and an obstacle, e.g., a cow, is directly before the signal. The signal signals ``Halt''. ATO GoA 4 adds the rule that a train halts before any detected obstacle. 

Trace links are able to enhance validation through simulation by tracing certain execution steps back to the original procedures.
Simulating the scenario shows that the train detects the cow and halts until the cow leaves the track, even if the signal switches to ``Go''.
Similarly, if the cow leaves and the signal is still signaling ``Halt'', the train waits.
At each point, the simulation displays the decisions of the ATO algorithms, e.g., if the obstacle is not detected anymore it is displayed why the system decided to halt. 

This allows us to check that rules for ATO GoA 4 do not override rules for normal train driving or otherwise interfere with them.
From a development process view, the simulation itself is a behavioral test that links its output with the requirements and the annotations are links for requirements traces.


\COMMENT{The example is available as a videos in the supplementary material at \url{formbar.raillab.de/videos19}.}
\subsection{Prototyping Rule Changes}
To reason about the effects of a proposed rule change, \visbar was first used to model a variant of the infrastructure in the west branch of the Frankfurt City Tunnel.
This branch has a length of 4.7km and is the main part of the Rhein-Main S-Bahn -- eight lines pass through it, with intervals below five minutes. 
Its high usage makes it representative of how rules affect operations in networks with high occupancy rate and short distances between signals.
We only model one direction (from Hauptbahnhof to S{\"u}dbahnhof) without the branch-off point Schlachthof, which is sufficient for the analyzed rule change.

On the infrastructure two trains with a 5 minutes interval are simulated, both with a maximal velocity of 60km/h. We model the following scenario: 
the main signal on the track between Ostendstra{\ss}e and Lokalbahnhof\footnote{This block was chosen because it has the shortest sight distance and requires the slowest speeds when driving on sight.} has a fault that is local to this signal (e.g., a broken bulb). 
To sustain operations, the train dispatcher gives an order to depart (equivalent to a Zs1 auxiliary signal) nonetheless.
\begin{description}
\item[Old rule.] The train dispatcher must not order to drive on sight, thus the first train can still drive the full 60km/h. In this case the second train, which departs at $t=300s$,  arrives in the final station at $t=1027s$.
\item[New rule.] Now, the train must first drive on sight, which is walking speed in tunnels (6 km/h). In this case the second train (which is not effected by this and may drive 60km/h) arrives at the final station at $t=1487s$.
\end{description}
The delay, over 7 minutes\footnote{This is longer than the delay caused purely by waiting (3 minutes) for the first train to arrive, but still realistic. The additional 4 minutes are caused by a non-optimal train dispatching in our model. 
However, the duties of the train dispatcher to document the situation and give written orders in case of faults accounts for this.} 
is specific to this infrastructure and time table, yet gives an estimate which helps the developers to assess the impact of a rule change. 
Another example to examine rule changes with \formbar is discussed in~\cite{scp}. \visbar is an improvement over the previous ad-hoc visualization, as it allows to assess the relevant information faster by showing the $v$-$t$ graph.

This application of \visbar was presented to the rulebook authors of DB Netz responsible for this rule, who deemed the visualization and the trace links as helpful.

\section{Conclusion and Future Research}\label{sec:conclusion}
We presented \visbar and have shown how it can be integrated into the processes for developing railway operation procedures.
It illustrates how model validation can be supported by integrating requirement traces and how these traces increase the usefulness of tests and visualization.
It extends our previous work on modeling these procedures by giving an interface that does not require the user to learn ABS to use the model, but gives him the possibility for deeper manipulation with ABS if necessary.
\visbar extends the use cases of formal tools in railway engineering from support for implementation and planning~\cite{ovado,dill} to the development of new procedures by using an ABS model to prototype ATO procedures.
Beyond railway engineering, we addressed the challenge to use technical documents as requirements for validation, which are not designed to be used as requirements and are not modifiable by the modeler.

\visbar is available under \url{formbar.raillab.de/visbar} with limited annotations, as most rulebooks and the rules for the above ATO case study are not public.
A video demonstrating the usage of \visbar is available under \url{https://figshare.com/s/71f1c2e7252bfd032f57}.

It is often observed that formal models offer a benefit for the designer, even without analyzing formal properties, as it forces to clarify all ambiguities. 
Thus, formal modeling languages must not only be easy to analyze, but also easy to validate and easy to integrate into existing development processes.
Yet, validation of formal system models and its place in development processes remains a challenging domain.
For future work, we are not only interested in the integration of validation of formal models into a development process, but the development process of formal models and digital twins itself.
In particular, we are interested in the following:
\begin{itemize}
\item Requirement trace generation for formal modeling. 
\item Integration of conceptual modeling~\cite{cmis} into formal model validation by connecting requirements and formal model with a domain ontology\footnote{Conceptual modeling faces similar problems with validation, but is more abstract in the information it captures and relies more on implicit knowledge, than specifications, designs or a concrete existing system.}.
\end{itemize}
The overarching questions are 
(1) how to design formal modeling languages (and IDEs for them) which are not only easily usable and analyzable, but also easy to validate
and (2) how to use traceability in verified formal models for certification.
We propose that automatic generation of traces would not only vastly simplify validation, but also be a step towards a wider acceptance of formal proofs for certification. 

\paragraph{Related Work}
Luteberget et al.~\cite{LutebergetCJS17} use traces to link errors raised during verification to the responsible part of the model and the original document. 
These traces roughly resemble the annotations in messages generated by \visbar during simulation, but are not used to validate the model itself. 
Ferrari et al.~\cite{FerrariGRTBFG18} investigated the requirements of railway engineering projects from a natural language processing perspective.
Concerning the connection of conceptual and formal modeling, Kharlamov et al.~\cite{semtwin} propose to use ontologies to develop digital twins, but not for validation. 

Fischer and Dghaym~\cite{FischerD19} use acceptance tests to validate a formal model of Hybrid ETCS L3 segments. 
Contrary to requirement traces and interactive visualization their approach requires \emph{fully formalized} test cases of observable behavior of the model.
This approach is not only subsumed by simulation -- as discussed, it also does not lower the cognitive burden of validation, as these test cases are a \emph{formal behavioral model themselves}.

Integration of multiple aspects is common for programming languages in mainstream IDEs, but 
development environments based on formal methods focus mostly only on the formal model and its verification, e.g., by an interface to the proof system.
E.g., the B-OVADO~\cite{ovado} tool for the PERF~\cite{perf} approach, offers a toolbox for data validation tasks that integrates B as a language to specify data.
The \texttt{Sphinx} tool~\cite{Mitsch}, which integrates verification and modeling tools for model-based engineering of hybrid systems, is the only approach that uses formal methods for coordinating multiple components for development.
It also provides a way to connect to documentation in a special UML profile and is specific to differential dynamic logic.
Interaction~\cite{Ladenberger17} and visualization~\cite{laden2} for validation of B-models was investigated by Ladenberger et al.
(Interactive) visualization of formal models is also supported for Circus~\cite{circus} and PVS~\cite{pvs} models.

\paragraph{Future Work} 
Beyond further research in the connection to requirements engineering sketched above, we plan (1) to enable statistical analyses, such as expected lost units~\cite{LU} after a rule change, in a representative network and (2) to integrate our verification approach~\cite{KamburjanH17} to use it for certifications.
We also plan to investigate how, analogous to Domain Specific Languages, Domain Specific IDEs, can be used to integrate formal methods into other domains. 

\FULL{
\paragraph{Acknowledgments} 
    This work is supported by the \formbar project, part of AG Signalling/DB RailLab.
    We thank Heike Villioth-Ebert, Armin Krieger, Matthias Kopitzki and Bilal {\"U}y{\"u}mez for their feedback.
}{}
\bibliographystyle{eptcs} 
\bibliography{refs}

\begin{thebibliography}{10}
\providecommand{\bibitemdeclare}[2]{}
\providecommand{\surnamestart}{}
\providecommand{\surnameend}{}
\providecommand{\urlprefix}{Available at }
\providecommand{\url}[1]{\texttt{#1}}
\providecommand{\href}[2]{\texttt{#2}}
\providecommand{\urlalt}[2]{\href{#1}{#2}}
\providecommand{\doi}[1]{doi:\urlalt{http://dx.doi.org/#1}{#1}}
\providecommand{\bibinfo}[2]{#2}

\bibitemdeclare{article}{barlas}
\bibitem{barlas}
\bibinfo{author}{Yaman \surnamestart Barlas\surnameend} (\bibinfo{year}{1996}):
  \emph{\bibinfo{title}{Formal Aspects of Model Validity and Validation in
  System Dynamics}}.
\newblock {\sl \bibinfo{journal}{System Dynamics Review - SYST DYNAM REV}}
  \bibinfo{volume}{12},
  \doi{10.1002/(SICI)1099-1727(199623)12:3<183::AID-SDR103>3.0.CO;2-4}.

\bibitemdeclare{inproceedings}{circus}
\bibitem{circus}
\bibinfo{author}{S.~L.~M. \surnamestart Barrocas\surnameend} \&
  \bibinfo{author}{Marcel \surnamestart Oliveira\surnameend}
  (\bibinfo{year}{2012}): \emph{\bibinfo{title}{JCircus 2.0: an Extension of an
  Automatic Translator from Circus to Java}}.
\newblock In \bibinfo{editor}{Peter~H. \surnamestart Welch\surnameend},
  \bibinfo{editor}{Frederick R.~M. \surnamestart Barnes\surnameend},
  \bibinfo{editor}{Kevin \surnamestart Chalmers\surnameend},
  \bibinfo{editor}{Jan~B{\ae}kgaard \surnamestart Pedersen\surnameend} \&
  \bibinfo{editor}{Adam~T. \surnamestart Sampson\surnameend}, editors: {\sl
  \bibinfo{booktitle}{34th Communicating Process Architectures, {CPA} 2012,
  organised under the auspices of WoTUG}}, \bibinfo{publisher}{Open Channel
  Publishing Ltd.}, pp. \bibinfo{pages}{15--36}.
\newblock \urlprefix\url{http://wotug.org/paperdb/show\_pap.php?f=1\&num=662}.

\bibitemdeclare{inproceedings}{perf}
\bibitem{perf}
\bibinfo{author}{Nazim \surnamestart Benaissa\surnameend},
  \bibinfo{author}{David \surnamestart Bonvoisin\surnameend},
  \bibinfo{author}{Abderrahmane \surnamestart Feliachi\surnameend} \&
  \bibinfo{author}{Julien \surnamestart Ordioni\surnameend}
  (\bibinfo{year}{2016}): \emph{\bibinfo{title}{The {PERF} Approach for Formal
  Verification}}.
\newblock In \bibinfo{editor}{Thierry \surnamestart Lecomte\surnameend},
  \bibinfo{editor}{Ralf \surnamestart Pinger\surnameend} \&
  \bibinfo{editor}{Alexander \surnamestart Romanovsky\surnameend}, editors:
  {\sl \bibinfo{booktitle}{{RSSRail 2016 proc.}}}, \bibinfo{publisher}{Springer
  International Publishing}, \bibinfo{address}{Cham}, pp.
  \bibinfo{pages}{203--214}, \doi{10.1007/978-3-319-33951-1\_15}.

\bibitemdeclare{article}{time}
\bibitem{time}
\bibinfo{author}{Joakim \surnamestart Bj{\o}rk\surnameend},
  \bibinfo{author}{Frank~S. \surnamestart de~Boer\surnameend},
  \bibinfo{author}{Einar~Broch \surnamestart Johnsen\surnameend},
  \bibinfo{author}{Rudolf \surnamestart Schlatte\surnameend} \&
  \bibinfo{author}{Silvia~Lizeth \surnamestart {Tapia Tarifa}\surnameend}
  (\bibinfo{year}{2013}): \emph{\bibinfo{title}{User-defined schedulers for
  real-time concurrent objects}}.
\newblock {\sl \bibinfo{journal}{{ISSE}}}
  \bibinfo{volume}{9}(\bibinfo{number}{1}), pp. \bibinfo{pages}{29--43},
  \doi{10.1007/s11334-012-0184-5}.

\bibitemdeclare{inproceedings}{isola}
\bibitem{isola}
\bibinfo{author}{Manfred \surnamestart Broy\surnameend}, \bibinfo{author}{Klaus
  \surnamestart Havelund\surnameend}, \bibinfo{author}{Rahul \surnamestart
  Kumar\surnameend} \& \bibinfo{author}{Bernhard \surnamestart
  Steffen\surnameend} (\bibinfo{year}{2018}): \emph{\bibinfo{title}{Towards a
  Unified View of Modeling and Programming (Track Introduction)}}.
\newblock In \bibinfo{editor}{Tiziana \surnamestart Margaria\surnameend} \&
  \bibinfo{editor}{Bernhard \surnamestart Steffen\surnameend}, editors: {\sl
  \bibinfo{booktitle}{{ISoLA}}}, \bibinfo{publisher}{Springer}, pp.
  \bibinfo{pages}{3--21}, \doi{10.1007/978-3-030-03418-4\_1}.

\bibitemdeclare{misc}{en}
\bibitem{en}
\bibinfo{author}{\surnamestart {CENELEC}\surnameend} (\bibinfo{year}{2011}):
  \emph{\bibinfo{title}{{DIN EN 50128:2011, Railway applications --
  Communication, Signalling and Processing Signals}}}.

\bibitemdeclare{misc}{ril408}
\bibitem{ril408}
\bibinfo{author}{\surnamestart {DB Netz AG, Frankfurt, Germany}\surnameend}
  (\bibinfo{year}{2017}): \emph{\bibinfo{title}{{Richtlinie 408,
  Fahrdienstvorschrift}}}.

\bibitemdeclare{misc}{ril819}
\bibitem{ril819}
\bibinfo{author}{\surnamestart {DB Netz AG, Frankfurt, Germany}\surnameend}
  (\bibinfo{year}{2017}): \emph{\bibinfo{title}{{Richtlinie 819, LST-Anlagen
  planen}}}.

\bibitemdeclare{inproceedings}{dill}
\bibitem{dill}
\bibinfo{author}{Stefan \surnamestart Dillmann\surnameend} \&
  \bibinfo{author}{Reiner \surnamestart H{\"{a}}hnle\surnameend}
  (\bibinfo{year}{2019}): \emph{\bibinfo{title}{Automated Planning of {ETCS}
  Tracks}}.
\newblock In: {\sl \bibinfo{booktitle}{RSSRail}}, {\sl \bibinfo{series}{LNCS}}
  \bibinfo{volume}{11495}, \bibinfo{publisher}{Springer}, pp.
  \bibinfo{pages}{79--90}, \doi{10.1007/978-3-030-18744-6\_5}.

\bibitemdeclare{misc}{eba}
\bibitem{eba}
\bibinfo{author}{\surnamestart {Eisenbahnbundesamt (Federal Railway
  Authority)}\surnameend} (\bibinfo{year}{2017}):
  \emph{\bibinfo{title}{Eisenbahn-Bau- und Betriebsordnung}}.
\newblock \bibinfo{note}{April 2017:
  https://www.gesetze-im-internet.de/ebo/index.html}.

\bibitemdeclare{article}{FerrariGRTBFG18}
\bibitem{FerrariGRTBFG18}
\bibinfo{author}{Alessio \surnamestart Ferrari\surnameend},
  \bibinfo{author}{Gloria \surnamestart Gori\surnameend},
  \bibinfo{author}{Benedetta \surnamestart Rosadini\surnameend},
  \bibinfo{author}{Iacopo \surnamestart Trotta\surnameend},
  \bibinfo{author}{Stefano \surnamestart Bacherini\surnameend},
  \bibinfo{author}{Alessandro \surnamestart Fantechi\surnameend} \&
  \bibinfo{author}{Stefania \surnamestart Gnesi\surnameend}
  (\bibinfo{year}{2018}): \emph{\bibinfo{title}{Detecting requirements defects
  with {NLP} patterns: an industrial experience in the railway domain}}.
\newblock {\sl \bibinfo{journal}{Empirical Software Engineering}}
  \bibinfo{volume}{23}(\bibinfo{number}{6}), pp. \bibinfo{pages}{3684--3733},
  \doi{10.1007/s10664-018-9596-7}.

\bibitemdeclare{inproceedings}{FischerD19}
\bibitem{FischerD19}
\bibinfo{author}{Tomas \surnamestart Fischer\surnameend} \&
  \bibinfo{author}{Dana \surnamestart Dghaym\surnameend}
  (\bibinfo{year}{2019}): \emph{\bibinfo{title}{Formal Model Validation Through
  Acceptance Tests}}.
\newblock In: {\sl \bibinfo{booktitle}{RSSRail 2019}}, {\sl
  \bibinfo{series}{LNCS}} \bibinfo{volume}{11495},
  \bibinfo{publisher}{Springer}, pp. \bibinfo{pages}{159--169},
  \doi{10.1007/978-3-030-18744-6\_10}.

\bibitemdeclare{inproceedings}{ovado}
\bibitem{ovado}
\bibinfo{author}{Manel \surnamestart Fredj\surnameend}, \bibinfo{author}{Sven
  \surnamestart Leger\surnameend}, \bibinfo{author}{Abderrahmane \surnamestart
  Feliachi\surnameend} \& \bibinfo{author}{Julien \surnamestart
  Ordioni\surnameend} (\bibinfo{year}{2017}): \emph{\bibinfo{title}{{OVADO} -
  Enhancing Data Validation for Safety-Critical Railway Systems}}.
\newblock In \bibinfo{editor}{Alessandro \surnamestart Fantechi\surnameend},
  \bibinfo{editor}{Thierry \surnamestart Lecomte\surnameend} \&
  \bibinfo{editor}{Alexander~B. \surnamestart Romanovsky\surnameend}, editors:
  {\sl \bibinfo{booktitle}{{RSSRail 2017 proc.}}}, {\sl \bibinfo{series}{LNCS}}
  \bibinfo{volume}{10598}, \bibinfo{publisher}{Springer}, pp.
  \bibinfo{pages}{87--98}, \doi{10.1007/978-3-319-68499-4\_6}.

\bibitemdeclare{article}{cosim}
\bibitem{cosim}
\bibinfo{author}{Cl{\'{a}}udio \surnamestart Gomes\surnameend},
  \bibinfo{author}{Casper \surnamestart Thule\surnameend},
  \bibinfo{author}{David \surnamestart Broman\surnameend},
  \bibinfo{author}{Peter~Gorm \surnamestart Larsen\surnameend} \&
  \bibinfo{author}{Hans \surnamestart Vangheluwe\surnameend}
  (\bibinfo{year}{2018}): \emph{\bibinfo{title}{Co-Simulation: {A} Survey}}.
\newblock {\sl \bibinfo{journal}{{ACM} Comput. Surv.}}
  \bibinfo{volume}{51}(\bibinfo{number}{3}), pp. \bibinfo{pages}{49:1--49:33},
  \doi{10.1145/3179993}.

\bibitemdeclare{article}{IEEE830}
\bibitem{IEEE830}
\bibinfo{author}{\surnamestart {IEEE}\surnameend} (\bibinfo{year}{1998}):
  \emph{\bibinfo{title}{{IEEE} Guide for Software Requirements
  Specifications}}.
\newblock {\sl \bibinfo{journal}{IEEE Std 830-1998}}.

\bibitemdeclare{inproceedings}{abs}
\bibitem{abs}
\bibinfo{author}{Einar~Broch \surnamestart Johnsen\surnameend},
  \bibinfo{author}{Reiner \surnamestart H{\"{a}}hnle\surnameend},
  \bibinfo{author}{Jan \surnamestart Sch{\"{a}}fer\surnameend},
  \bibinfo{author}{Rudolf \surnamestart Schlatte\surnameend} \&
  \bibinfo{author}{Martin \surnamestart Steffen\surnameend}
  (\bibinfo{year}{2010}): \emph{\bibinfo{title}{{ABS:} {A} Core Language for
  Abstract Behavioral Specification}}.
\newblock In: {\sl \bibinfo{booktitle}{{FMCO}}}, {\sl \bibinfo{series}{LNCS}}
  \bibinfo{volume}{6957}, \bibinfo{publisher}{Springer},
  \doi{10.1007/978-3-642-25271-6\_8}.

\bibitemdeclare{inproceedings}{KamburjanH17}
\bibitem{KamburjanH17}
\bibinfo{author}{Eduard \surnamestart Kamburjan\surnameend} \&
  \bibinfo{author}{Reiner \surnamestart H{\"{a}}hnle\surnameend}
  (\bibinfo{year}{2017}): \emph{\bibinfo{title}{Deductive Verification of
  Railway Operations}}.
\newblock In: {\sl \bibinfo{booktitle}{RSSRail 2017}}, {\sl
  \bibinfo{series}{LNCS}} \bibinfo{volume}{10598},
  \bibinfo{publisher}{Springer}, pp. \bibinfo{pages}{131--147},
  \doi{10.1007/978-3-319-68499-4\_9}.

\bibitemdeclare{inproceedings}{proto}
\bibitem{proto}
\bibinfo{author}{Eduard \surnamestart Kamburjan\surnameend} \&
  \bibinfo{author}{Reiner \surnamestart H\"ahnle\surnameend}
  (\bibinfo{year}{2018}): \emph{\bibinfo{title}{Prototyping Formal System
  Models with Active Objects}}.
\newblock In: {\sl \bibinfo{booktitle}{Interaction and Concurrency
  Experience}}, {\sl \bibinfo{series}{EPTCS}} \bibinfo{volume}{279},
  \bibinfo{publisher}{Open Publishing Association}, pp.
  \bibinfo{pages}{52--67}, \doi{10.4204/EPTCS.279.7}.

\bibitemdeclare{article}{scp}
\bibitem{scp}
\bibinfo{author}{Eduard \surnamestart Kamburjan\surnameend},
  \bibinfo{author}{Reiner \surnamestart H{\"a}hnle\surnameend} \&
  \bibinfo{author}{Sebastian \surnamestart Sch{\"o}n\surnameend}
  (\bibinfo{year}{2018}): \emph{\bibinfo{title}{Formal modeling and analysis of
  railway operations with active objects}}.
\newblock {\sl \bibinfo{journal}{Science of Computer Programming}}
  \bibinfo{volume}{166}, pp. \bibinfo{pages}{167 -- 193},
  \doi{10.1016/j.scico.2018.07.001}.

\bibitemdeclare{mastersthesis}{LU}
\bibitem{LU}
\bibinfo{author}{Florian~Rudolf \surnamestart K{\"a}mmerer\surnameend}
  (\bibinfo{year}{2017}): \emph{\bibinfo{title}{{Entwicklung eines
  Kennzahlensystems f{\"u}r Effektivit{\"a}t des Bahnbetriebs bei Abweichungen
  vom Regelbetrieb}}}.
\newblock Master's thesis, \bibinfo{school}{Technische Universit{\"a}t
  Darmstadt}.

\bibitemdeclare{inproceedings}{semtwin}
\bibitem{semtwin}
\bibinfo{author}{E.~\surnamestart {Kharlamov}\surnameend},
  \bibinfo{author}{F.~\surnamestart {Martin-Recuerda}\surnameend},
  \bibinfo{author}{B.~\surnamestart {Perry}\surnameend},
  \bibinfo{author}{D.~\surnamestart {Cameron}\surnameend},
  \bibinfo{author}{R.~\surnamestart {Fjellheim}\surnameend} \&
  \bibinfo{author}{A.~\surnamestart {Waaler}\surnameend}
  (\bibinfo{year}{2018}): \emph{\bibinfo{title}{Towards Semantically Enhanced
  Digital Twins}}.
\newblock In: {\sl \bibinfo{booktitle}{2018 IEEE International Conference on
  Big Data}}, pp. \bibinfo{pages}{4189--4193},
  \doi{10.1109/BigData.2018.8622503}.

\bibitemdeclare{phdthesis}{Ladenberger17}
\bibitem{Ladenberger17}
\bibinfo{author}{Lukas \surnamestart Ladenberger\surnameend}
  (\bibinfo{year}{2017}): \emph{\bibinfo{title}{Rapid Creation of Interactive
  Formal Prototypes for Validating Safety-Critical Systems}}.
\newblock Ph.D. thesis, \bibinfo{school}{University of D{\"{u}}sseldorf,
  Germany}.

\bibitemdeclare{inproceedings}{laden2}
\bibitem{laden2}
\bibinfo{author}{Lukas \surnamestart Ladenberger\surnameend},
  \bibinfo{author}{Jens \surnamestart Bendisposto\surnameend} \&
  \bibinfo{author}{Michael \surnamestart Leuschel\surnameend}
  (\bibinfo{year}{2009}): \emph{\bibinfo{title}{Visualising Event-B Models with
  B-Motion Studio}}.
\newblock In \bibinfo{editor}{Mar{\'i}a \surnamestart Alpuente\surnameend},
  \bibinfo{editor}{Byron \surnamestart Cook\surnameend} \&
  \bibinfo{editor}{Christophe \surnamestart Joubert\surnameend}, editors: {\sl
  \bibinfo{booktitle}{Formal Methods for Industrial Critical Systems}},
  \bibinfo{publisher}{Springer Berlin Heidelberg}, \bibinfo{address}{Berlin,
  Heidelberg}, pp. \bibinfo{pages}{202--204},
  \doi{10.1007/978-3-642-04570-7\_17}.

\bibitemdeclare{inproceedings}{LutebergetCJS17}
\bibitem{LutebergetCJS17}
\bibinfo{author}{Bj{\o}rnar \surnamestart Luteberget\surnameend},
  \bibinfo{author}{John~J. \surnamestart Camilleri\surnameend},
  \bibinfo{author}{Christian \surnamestart Johansen\surnameend} \&
  \bibinfo{author}{Gerardo \surnamestart Schneider\surnameend}
  (\bibinfo{year}{2017}): \emph{\bibinfo{title}{Participatory Verification of
  Railway Infrastructure by Representing Regulations in RailCNL}}.
\newblock In \bibinfo{editor}{Alessandro \surnamestart Cimatti\surnameend} \&
  \bibinfo{editor}{Marjan \surnamestart Sirjani\surnameend}, editors: {\sl
  \bibinfo{booktitle}{Software Engineering and Formal Methods - 15th
  International Conference, {SEFM} 2017, Trento, Italy, September 4-8, 2017,
  Proceedings}}, {\sl \bibinfo{series}{LNCS}} \bibinfo{volume}{10469},
  \bibinfo{publisher}{Springer}, pp. \bibinfo{pages}{87--103},
  \doi{10.1007/978-3-319-66197-1\_6}.

\bibitemdeclare{article}{Mitsch}
\bibitem{Mitsch}
\bibinfo{author}{Stefan \surnamestart Mitsch\surnameend},
  \bibinfo{author}{Grant~Olney \surnamestart Passmore\surnameend} \&
  \bibinfo{author}{Andr{\'{e}} \surnamestart Platzer\surnameend}
  (\bibinfo{year}{2014}): \emph{\bibinfo{title}{Collaborative
  Verification-Driven Engineering of Hybrid Systems}}.
\newblock {\sl \bibinfo{journal}{Mathematics in Computer Science}}
  \bibinfo{volume}{8}(\bibinfo{number}{1}), pp. \bibinfo{pages}{71--97},
  \doi{10.1007/s11786-014-0176-y}.

\bibitemdeclare{inproceedings}{tla}
\bibitem{tla}
\bibinfo{author}{Chris \surnamestart Newcombe\surnameend}
  (\bibinfo{year}{2014}): \emph{\bibinfo{title}{Why {Amazon} Chose
  {TLA}{\thinspace}+{\thinspace}}}.
\newblock In \bibinfo{editor}{Yamine \surnamestart Ait~Ameur\surnameend} \&
  \bibinfo{editor}{Klaus-Dieter \surnamestart Schewe\surnameend}, editors: {\sl
  \bibinfo{booktitle}{Abstract State Machines, Alloy, B, TLA, VDM, and Z}},
  \bibinfo{publisher}{Springer Berlin Heidelberg}, \bibinfo{address}{Berlin,
  Heidelberg}, pp. \bibinfo{pages}{25--39}, \doi{10.1007/978-3-662-43652-3\_3}.

\bibitemdeclare{misc}{bdd}
\bibitem{bdd}
\bibinfo{author}{Dan \surnamestart North\surnameend} (\bibinfo{year}{2006}):
  \emph{\bibinfo{title}{Introducing {BDD}}}.
\newblock \bibinfo{note}{Http://dannorth.net/introducing-bdd/}.

\bibitemdeclare{book}{cmis}
\bibitem{cmis}
\bibinfo{author}{Antoni \surnamestart Oliv{\'e}\surnameend}
  (\bibinfo{year}{2007}): \emph{\bibinfo{title}{Conceptual Modeling of
  Information Systems}}.
\newblock \bibinfo{publisher}{Springer-Verlag}, \bibinfo{address}{Berlin,
  Heidelberg}.

\bibitemdeclare{article}{pachlBahn}
\bibitem{pachlBahn}
\bibinfo{author}{Jörn \surnamestart Pachl\surnameend} (\bibinfo{year}{2018}):
  \emph{\bibinfo{title}{{Das Ersatzsignal -- ein deutscher Sonderweg?}}}
\newblock {\sl \bibinfo{journal}{{Deine Bahn}}} \bibinfo{volume}{3}.
\newblock \bibinfo{note}{{In German}}.

\bibitemdeclare{article}{twin}
\bibitem{twin}
\bibinfo{author}{Roland \surnamestart Rosen\surnameend}, \bibinfo{author}{Georg
  \surnamestart von Wichert\surnameend}, \bibinfo{author}{George \surnamestart
  Lo\surnameend} \& \bibinfo{author}{Kurt~D. \surnamestart
  Bettenhausen\surnameend} (\bibinfo{year}{2015}): \emph{\bibinfo{title}{About
  The Importance of Autonomy and Digital Twins for the Future of
  Manufacturing}}.
\newblock {\sl \bibinfo{journal}{IFAC-PapersOnLine}}
  \bibinfo{volume}{48}(\bibinfo{number}{3}), pp. \bibinfo{pages}{567 -- 572},
  \doi{10.1016/j.ifacol.2015.06.141}.

\bibitemdeclare{misc}{do}
\bibitem{do}
\bibinfo{author}{\surnamestart {RTCA Inc, EUROCAE}\surnameend}
  (\bibinfo{year}{2012}): \emph{\bibinfo{title}{{DO-178C}}}.

\bibitemdeclare{inbook}{arbab}
\bibitem{arbab}
\bibinfo{author}{Rudolf \surnamestart Schlatte\surnameend},
  \bibinfo{author}{Einar~Broch \surnamestart Johnsen\surnameend},
  \bibinfo{author}{Jacopo \surnamestart Mauro\surnameend},
  \bibinfo{author}{Silvia~Lizeth \surnamestart {Tapia Tarifa}\surnameend} \&
  \bibinfo{author}{Ingrid~Chieh \surnamestart Yu\surnameend}
  (\bibinfo{year}{2018}): \emph{\bibinfo{title}{Release the Beasts: When Formal
  Methods Meet Real World Data}}, pp. \bibinfo{pages}{107--121}.
\newblock \bibinfo{publisher}{Springer International Publishing},
  \bibinfo{address}{Cham}, \doi{10.1007/978-3-319-90089-6_8}.

\bibitemdeclare{article}{Bilal1}
\bibitem{Bilal1}
\bibinfo{author}{Bilal \surnamestart {\"U}y{\"u}mez\surnameend}
  (\bibinfo{year}{2018}): \emph{\bibinfo{title}{{Modellierung des
  Steuerungsprozesses der R{\"u}ckfallebenen als Grundlage f{\"u}r die
  Automatisierung}}}.
\newblock {\sl \bibinfo{journal}{Eisenbahntechnische Rundschau}}.
\newblock \bibinfo{note}{In German}.

\bibitemdeclare{inproceedings}{pvs}
\bibitem{pvs}
\bibinfo{author}{Nathaniel \surnamestart Watson\surnameend},
  \bibinfo{author}{Steve \surnamestart Reeves\surnameend} \&
  \bibinfo{author}{Paolo \surnamestart Masci\surnameend}
  (\bibinfo{year}{2018}): \emph{\bibinfo{title}{Integrating User Design and
  Formal Models within PVSio-Web}}.
\newblock In \bibinfo{editor}{Paolo \surnamestart Masci\surnameend},
  \bibinfo{editor}{Rosemary \surnamestart Monahan\surnameend} \&
  \bibinfo{editor}{Virgile \surnamestart Prevosto\surnameend}, editors: {\sl
  \bibinfo{booktitle}{Proceedings 4th Workshop on Formal Integrated Development
  Environment, F-IDE@FLoC 2018, Oxford, England, 14 July 2018.}}, {\sl
  \bibinfo{series}{{EPTCS}}} \bibinfo{volume}{284}, pp.
  \bibinfo{pages}{95--104}, \doi{10.4204/EPTCS.284.8}.

\end{thebibliography}
\clearpage
%
\end{document}